\documentclass[10pt,letterpaper]{article}
 \usepackage{opex3_arXiv}
 \usepackage{color}
\begin{document}

\title{Multi-pass oscillator layout for high-energy mode-locked thin-disk lasers}

\author{K. Schuhmann$^{1,2,*}$,  K. Kirch$^{1,2}$ and  A. Antognini$^{1,2}$ }

\address{$^1$ Institute for Particle Physics, ETH, 8093 Zurich, Switzerland\\
$^2$ Paul Scherrer Institute, 5232 Villigen-PSI, Switzerland}

\email{$^*$ skarsten@phys.ethz.ch} 



\begin{abstract}

A novel optical layout for a multi-pass resonator is presented paving
the way for pulse energy scaling of mode-locked thin-disk lasers.
The multi-pass resonator we are proposing consists of a concatenation
of nearly identical optical segments.
Each segment corresponds to a round-trip in an optically stable cavity
containing an active medium exhibiting soft aperture effects.
This scheme is apt for energy and power scaling because the stability
region of this multi-pass resonator contrarily to the 4f-based schemes
does not shrink with the number of passes.
Simulation of the eigen-mode of this multi-segment resonator requires
considering aperture effects.
This has been achieved by implementing effective Gaussian apertures
into the ABCD-matrix formalism as lenses with
imaginary focal length.
We conclude proposing a simple way to double the stability region of
the state-of-the-art layouts used in industry achievable by a
minimal rearrangement of the used optical components.
\end{abstract}



\bibliographystyle{osajnl}





\section{Motivation}

Ultra-short laser pulse sources~\cite{Keller2010, Krausz2009} enable a
large variety of fundamental physics investigations, as well as
technological and industrial applications.
Many applications in industry and strong-field physics will
tremendously benefit from the increase of the pulse energy in the mJ
regime at few MHz repetition rates~\cite{Sudmeyer2008}: on one hand
production throughput and material compendium extension especially for
materials where non-linear multi-photon absorption is required, and on
the other hand, reduced measurement times, increased signal to noise,
and new scientific possibilities.

Mode-locked thin-disk~\cite{Giesen1994, Giesen2007a} lasers are
widely used in research laboratories and in industry because of
their power scaling and high pulse energy
capabilities~\cite{TRUMPF_www, Gottwald2012, Piehler2012, Mende2009,
  Fattahi2014, Larionov2014, Saraceno2015, Brons2014, Tummler2009,
  Negel2015, Kanda2013, Pronin2012}.
The output pulse energy $E$ of a mode-locked thin-disk laser can be increased,
at a given average output power $P_\mathrm{avg}$, by reducing the
laser repetition rate $f_\mathrm{rep}$, given the simple relation
$E=P_\mathrm{avg}/f_\mathrm{rep}$ from energy conservation.
Smaller repetition rates can be achieved simply by increasing the
oscillator cavity length. 
One successful way to increment the resonator length was obtained by
inserting into the cavity a Herriott cell~\cite{Richardson2009, Saraceno2014}.
However, the elevated intra-cavity pulse energy achieved in this way
required operation of the oscillator in an evacuated environment to
avoid detrimental non-linear effects in air~\cite{Saraceno2014}.

The cavity length can be also increased by folding the laser beam on
the active medium (thin-disk) several times per
round-trip~\cite{Gottwald2012, Bauer2012}.
The large gain per round-trip achievable with such an active
multi-pass cell enables large output coupling, which brings along a
reduction of the intra-cavity power.
Hence, this scheme providing a long cavity and decreased intra-cavity
intensity is twofold advantageous and is qualified for industrial
applications as it allows operation in air.
Another important feature of a multi-pass resonator scheme is the
reduction of Q-switching instabilities due to a linear decrease of the
gain saturation fluence with the number of reflections at the
thin-disk~\cite{Honninger1999, Dannecker2014}.

\begin{figure}[t!]
\centering\includegraphics[width=0.75\textwidth]{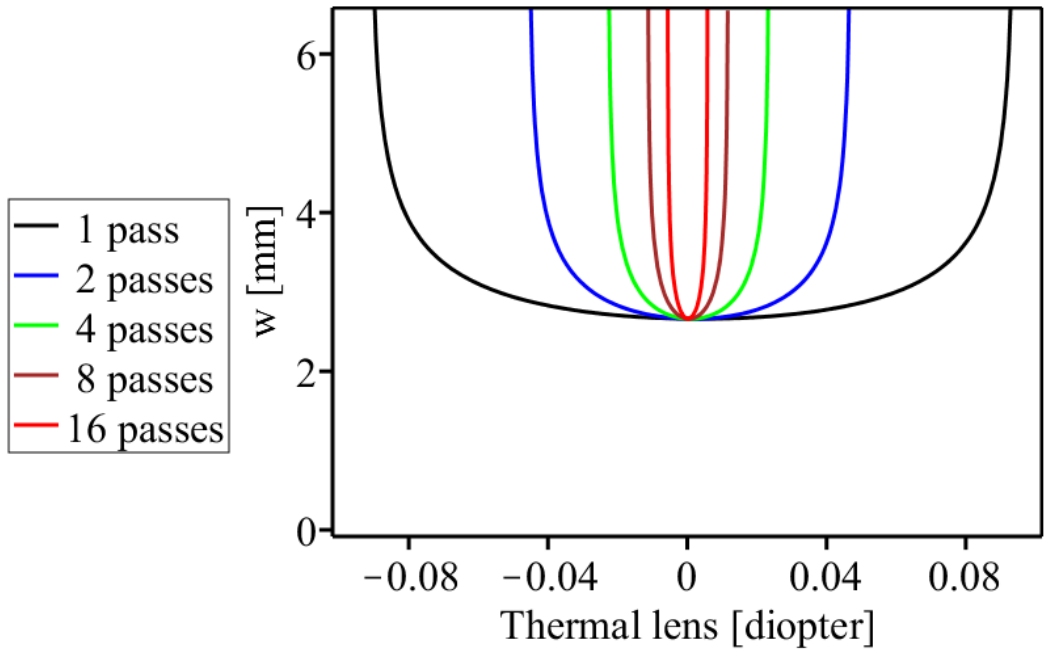}
\caption{\label{fig:1} (Color online) Stability plots of multi-pass
  resonators based on 4f-imaging stages.  Plotted are the cavity
  eigen-mode (TEM00-mode) waist w at the thin-disk position for
  variations of the thin-disk thermal lens from the layout value.  The
  shrinking of the stability region with the number of passes per
  cavity round-trip arising in 4f-based multi-pass resonators is
  demonstrated.  We computed these diagrams using the ABCD-matrix
  formalism and by embedding the 4f-stages at the thin-disk position
  of the stable resonator depicted in Fig.~\ref{fig:2}~(b). A
  wavelength of 1030~nm was used. For a given eigen-mode size the
  stability plots do not depend on the specific layout of the
  resonator.  }
\end{figure}
The multi-pass active cells realized to date~\cite{Gottwald2012,
  Bauer2012} are based on relay 4f-imaging: 4f optical segments are
used to image the thin-disk from pass to pass so that the beam
propagation in the active multi-pass cell proceeds following the
scheme disk-4f-disk-4f-disk-4f$\cdots$ .
The 4f propagation from the optical point of view corresponds to a
zero effective length propagation and it does not provide stability
for misalignment or variation of the thin-disk focal strength.
Hence, to realize a stable laser operation, the 4f multi-pass cell has
to be embedded in a stable optical resonator~\cite{Gottwald2012}.
The 4f multi-pass cell with $N$ number of passes can be described as a
single-pass having a total optical length of
$L_\mathrm{multi-pass}=(N-1)L$, a gain of $g_\mathrm{multi-pass}=g^N$,
and an active medium dioptric power of $V_\mathrm{multi-pass}=NV$,
where $L=4f$ represents the length of a single 4f-imaging stage, $g$ the
single-pass gain, and $V$ the thin-disk dioptric power.
Due to these cumulative effects, the resonator stability
zones~\cite{magni1986} of an oscillator containing such a multi-pass
4f-based cell shrink linearly with the number of passes $N$ as shown
in Fig.~\ref{fig:1} for variations of the disk thermal
lens~\cite{Speiser2009}.
This shrinking limits energy and power scaling~\cite{Baer2012}.

In summary, the 4f-based multi-pass oscillators show a limited
energy scaling (capitalizing only on the advantages related to the
long cavity length and the reduction of the intracavity circulating
intensity) but suffer for the shrinking of the stability region with
the number of passes which reduces the maximal achievable output
power.
%
%
In this paper a novel multi-pass resonator scheme is presented which overcomes
the thermal lens related power and energy limitations of
state-of-the-art multi-pass mode-locked laser oscillators.
%
In Sec.~\ref{sec:our-idea} our multi-pass scheme is exposed whose 
stability regions do not shrink with the number of passes. This opens
the way for further energy and power scaling.
A preliminary proof of principle of this new scheme is given in
Sec.~\ref{sec:proof} while in Sec.~\ref{sec:merging} a design merging
the to date 4f-based industrial scheme with our scheme is presented.

\begin{figure}[t!]
\centering\includegraphics[width=0.95\textwidth]{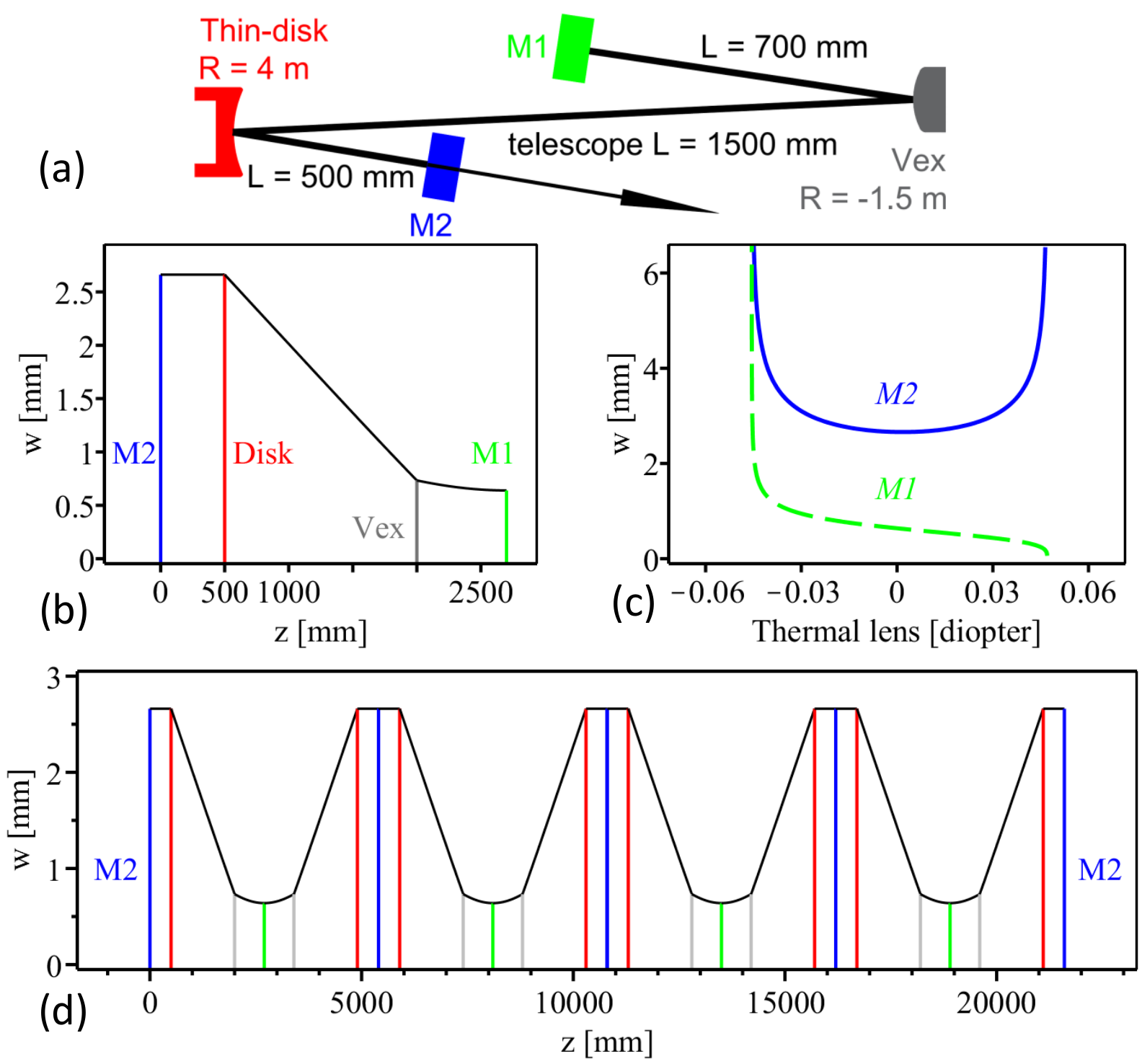}
\caption{\label{fig:2} (Color online) Scheme of the optical
  layout and properties of a multi-pass resonator as a
  concatenation of identical  optically stable segments. (a) Standard
  thin-disk laser resonator layout given by a flat end-mirror M1, a
  convex mirror, a concave thin-disk (red) and a flat end-mirror
  M2. (b) Corresponding eigen-mode waist w evolution along the
  resonator.  (c) Corresponding stability plot. Plotted is the
  eigen-mode waist at the M1 and M2 mirror positions for variations of
  the thin-disk thermal lens from the layout value. (d) Multi-pass
  resonator layout and eigen-mode waist evolution resulting by
  concatenating 8 times the optical segment of (a). The stability plot
  for the multi-segment resonator is identical to the single-segment
  stability plot shown in (c). The position of the optical elements
  are indicated by the vertical lines. The diagrams have been computed
  using the ABCD-matrix formalism and assuming TEM00 mode.}
\end{figure}

\section{New multi-pass resonator design}
\label{sec:our-idea}

The multi-pass resonator we are proposing is based on a concatenation
of identical (or nearly identical) segments. 
Each segment corresponds to a round-trip in an optically stable
resonator containing one pass (or more) on the same active medium,
which exhibits soft-aperture effects.

Since the multi-pass oscillator is inheriting the eigen-mode
properties of the underlying segment, we design this segment to be
stable and insensitive to thermal lens variations.
An example of a stable resonator whose round-trip propagation gives
rise to a segment is shown in Fig.~\ref{fig:2}~(a).
It is formed by a plane end-mirror M2, a thin-disk acting as concave
mirror, a convex mirror (Vex) and a flat end-mirror M1.
The eigen-mode waist $w$ evolution along this cavity is shown in
Fig.~\ref{fig:2}~(b).
This cavity  is widely used~\cite{magni1986} because it provides an
out-coupling mirror M2 with out-coupled beam waist (and divergence)
insensitive to variations of the thin-disk thermal lens as
demonstrated by the blue continuous curve in Fig.~\ref{fig:2}~(c)
representing the eigen-mode waist at the mirror M2 position for
variations of the thin-disk thermal lens.
For comparison the beam waist at the other end-mirror M1 which
features a larger dependency on the thermal lens variation is given as
well (green dashed curve).
This resonator layout is extensively used also because it allows for
simple adjustments of the mode properties: the distance between
thin-disk and the convex mirror can be adapted to shift the stability
region of the cavity, while the beam waist in the center of the
stability region can be adjusted by adapting the distance between the
M1 and the convex mirrors.

\begin{figure}[t!]
\centering\includegraphics[width=0.95\textwidth]{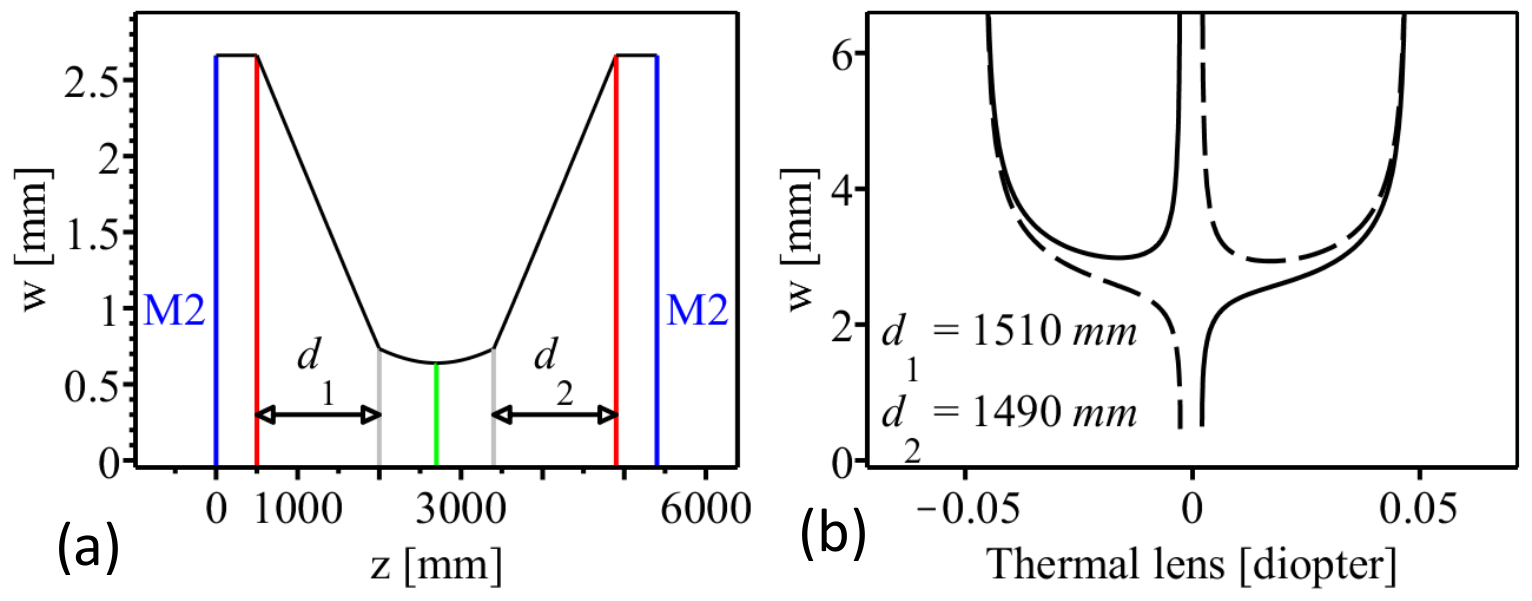}
\caption{\label{fig:3} (Color online) (a) Optical layout and
  eigen-mode waist w of a two-segment resonator with 4 reflections at
  the thin-disk per round-trip. (b) Corresponding stability plot.
  Plotted is the eigen-mode waist w at the two end-mirror positions M2
  for variations of the thin-disk thermal lens from the layout
  value. A small asymmetry has been introduced ($d_1\neq d_2$) between the two
  segments which induces a discontinuity  in the center
  of the stability region (cf. with Fig.~\ref{fig:2}~(c)).}
\end{figure}

As already mentioned, the multi-pass resonator according to our scheme
is obtained by concatenating multiple times the same optically stable
segment: each segment corresponding to a round-trip propagation in a
stable cavity.
An example of such a concatenation is shown in Fig.~\ref{fig:2}~(d)
where 8 segments based on the cavity shown in Fig.~\ref{fig:2}~(a) enables 16
reflections at the thin-disk per round-trip.
The stability regions of this multi-pass resonator coincide with the
stability regions of a single segment (given in Fig.~\ref{fig:2}~(c))
provided all segments are identical.
However, small differences between segments are unavoidable when
practically realizing a multi-segment oscillator
because of small variations of propagation lengths, incident angles
and mirror curvatures.

Consideration of this segment-to-segment asymmetries is essential for
the understanding of the here proposed new multi-pass resonator
concept.
In fact, this design has been discarded by the thin-disk laser
community because apparently these asymmetries prompt the formation
of gaps in the stability region.
The gap size depends on the extent of the segment-to-segment asymmetry.
The formation of these gaps as a consequence of small
segment-to-segment differences is exemplified in Fig.~\ref{fig:3} for
the particularly simple case that the multi-pass resonator is composed
of only two segments.
In the two-segment case, the gap arises in the center of the
stability region.
Similarly, for a multi-pass oscillator with several segments and
various segment-to-segment deviations, a multitude of disruptions
would apparently fragment and reduce the original stability region (of
the single segment).
It seems thus that the segment-to-segment asymmetries would undermine
the usefulness of this scheme.

The stability plots shown in Fig.~\ref{fig:1},~\ref{fig:2} and
\ref{fig:3} have been computed using the ABCD-matrix formalism.
This formalism is a powerful instrument to compute eigen-mode and
stability regions of resonators.
However, as already noted in~\cite{Gatz1994}, it is mostly used for
computing bare resonators neglecting the effect of the transversely
varying gain in the active material.
Aperture effects which naturally occur in a pumped active medium
mainly due to gain (absorption) in the pumped (unpumped) regions and
related diffraction (mainly outside the pumped spot) may significantly
affect the eigen-mode and stability properties of the
resonator~\cite{Casperon1975B}.
These effects can be described approximatively by a Gaussian aperture
at the active medium and included into the ABCD-matrix formalism as
imaginary lens~\cite{Casperon1975, Herrman1994,
  Kogelnik1965, Siegman1986}.
The ABCD-matrix describing the thin-disk  can
be thus written as
\begin{equation}
  M_\mathrm{thin-disk}= 
  \left[
  \begin{array}{ c c }
     1 & 0 \\
     -\frac{1}{f}-i\frac{\lambda}{\pi W^2} & 1
  \end{array} \right]\; ,
\label{eq:matrix}
\end{equation}
where $W$ represents the effective waist of the Gaussian aperture,
$\lambda$ the laser wavelength and $f$ the thin-disk focal length
which also includes thermal lens effects.
\begin{figure}[t!]
\centering\includegraphics[width=0.95\textwidth]{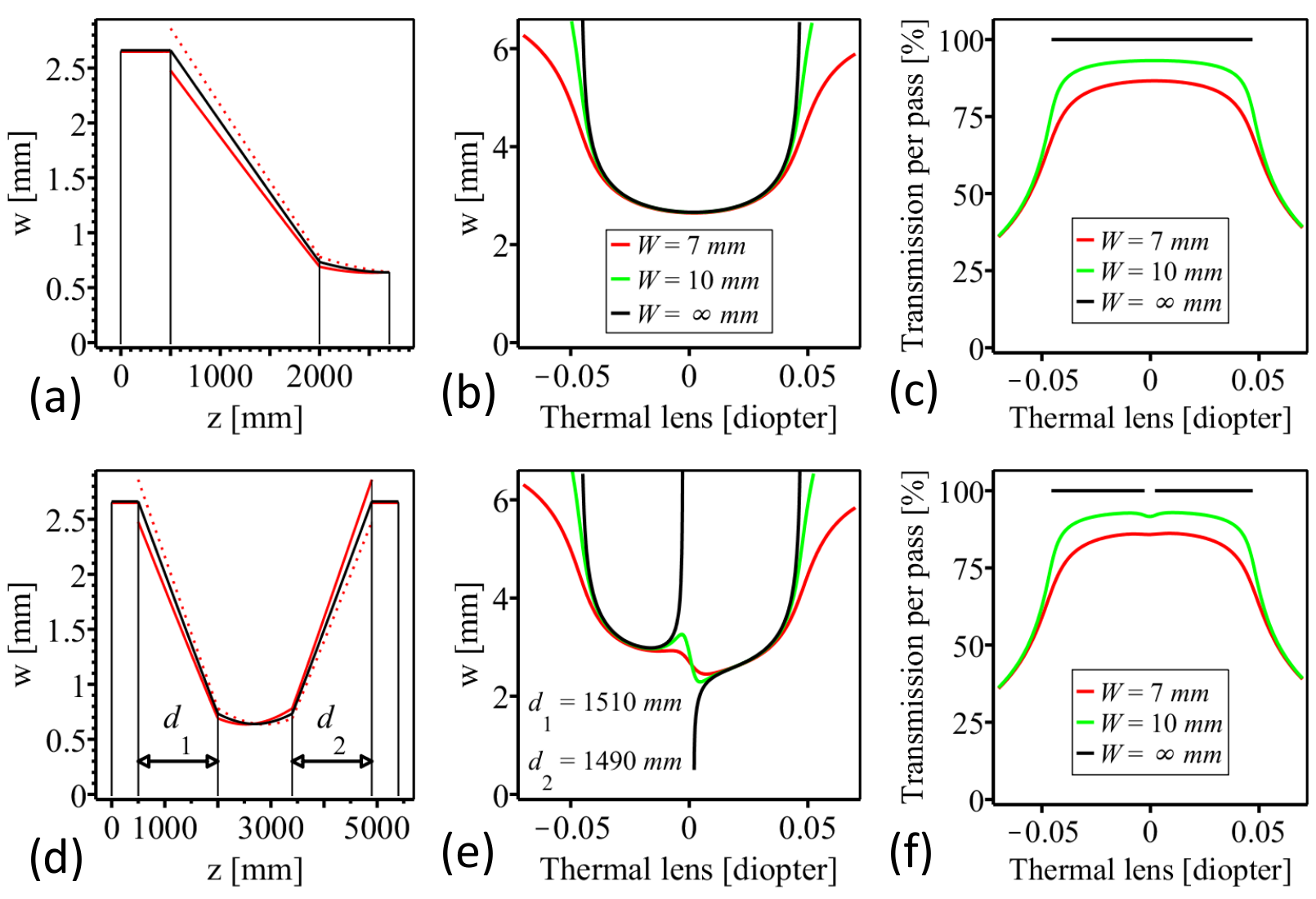}
\caption{\label{fig:4} (Color online) Influence of a Gaussian aperture
  at the active medium on the properties of a single-segment (top row)
  and a multi-segment (bottom row) resonators. (a) Optical layout and
  eigen-mode waist w evolution of a single-segment resonator. The
  black curve has been computed without aperture effects. The two red
  curves represent the back (dotted) and forth (continuous)
  propagation when aperture effects are included. (b) Corresponding
  stability plot. Plotted is the waist at the left end-mirror position
  for variations of the thin-disk thermal lens without aperture
  effects (black) and for two aperture waists $W$ (green and red). (c)
  Average (over a round-trip) transmission of the eigen-mode through
  the Gaussian aperture for variations of the thermal lens. For an
  infinite sized aperture the transmission (defined only within the
  stability region) is 100\%. (d) Similar to (a) but for a two-segment
  resonator. (e) Similar to (b) but for a two-segment resonator where
  a small asymmetry between the two segments has been introduced. The
  aperture effects damp the instability and close the gap in the
  stability region. (f) Similar to (c) for the two-segment resonator
  with the above specified small asymmetry. The increase of losses at
  the original gap position is minimal. All the curves have been computed
  using the ABCD-matrix formalism allowing for lenses with complex
  values. }
\end{figure}

Standard resonator designs do not include soft aperture effects
because for a single-segment resonator (see Fig.~\ref{fig:4}~(a)), the
inclusion of aperture effects does not considerably alter the computed
value of the eigen-mode size for thin-disk thermal lens variations
within the ``original'' (computed without considering the soft
aperture effect) stability region (see Fig.~\ref{fig:4}~(b)).
For dioptric power outside the ``original'' stability range, the
inclusion of aperture effects results in eigen-modes with finite waist
which implies an extension of the stability region~\cite{Gatz1994,
  Willetts1988}.
Therefore the inclusion of aperture effects shows that in
principle laser operation may occur also outside the ``original''
stability region.
Yet this extension has no practical relevance
because outside the ``original'' stability range the round-trip losses
caused by the aperture are increasing dramatically as demonstrated in
Fig.~\ref{fig:4}~(c) and~\cite{Gatz1994, Magni1996}.

Contrarily, aperture effects need to be included in the simulations of
multi-segment resonators (e.g. Fig.~\ref{fig:4}~(d)) having small
segment-to-segment deviations.
Simulating multi-segment resonators with small segment-to-segment
deviations without accounting for soft aperture effects produces 
wrong results because it predicts the formation of gaps within the
stability region which does not occur in reality.
The inclusion of these soft apertures into the simulations suppresses
these gaps as shown in Fig.~\ref{fig:4}~(e) for the particular case
of a two-segment resonator and leaves residual small fluctuations of
the eigen-mode waist.
Hence, it is essential to compute the stability properties of the
multi-pass resonator including soft aperture effects.
However, it is
important to stress that the general behavior of the stability regions
does not critically depend on the exact value of the assumed aperture waist
$W$  as can be deduced by comparing the green with the  red curves of
Fig.~\ref{fig:4}~(e).

Also the aperture-related losses per pass (averaged over a round-trip)
confirm that when including soft-aperture effects the multi-segment
resonator with small asymmetries behaves similar to the single-segment
resonator.
The residual waist fluctuations arising from the suppression of the gap
give rise to a negligible increase of losses per pass compared with
the single-segment case as visible by comparing Fig.~\ref{fig:4}~(f)
with Fig.~\ref{fig:4}~(c).
\begin{figure}[t!]
\centering\includegraphics[width=0.95\textwidth]{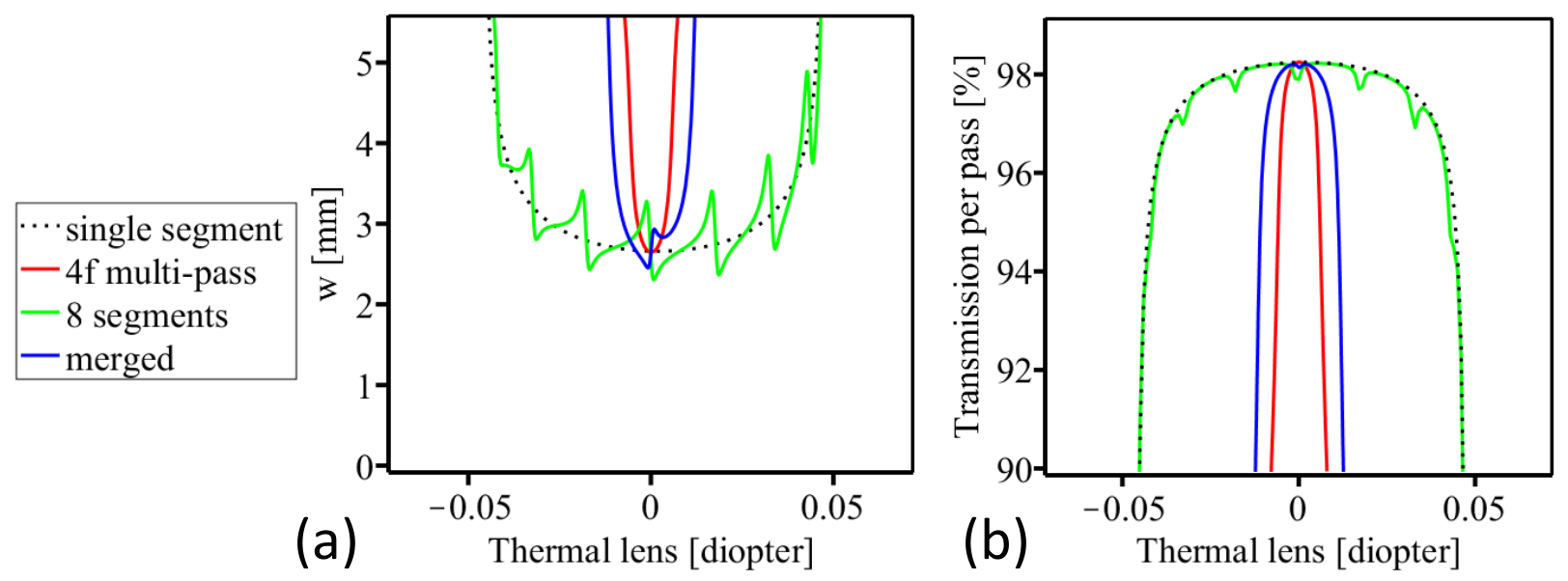}
\caption{\label{fig:5} (Color online) (a) Stability properties of
  three multi-pass resonator designs having same eigen-mode waist w,
  16 reflections (per round-trip) at the active medium and a gain
  medium with a Gaussian aperture of $W=20$~mm. The black dotted curve
  (1 segment containing two passes) represents the stability plot of
  the single-segment resonator of Fig.~\ref{fig:2}~(b).  It serves as
  reference.  The green curve (8 segments, each containing 2 passes)
  represents our design as a succession of nearly-identical segments
  as given in Fig.~\ref{fig:2}~(d). The fluctuations arise from a
  small segment-to-segment asymmetry: the distance disk to convex
  mirror in the first segment has been assumed to be 1518~mm i.e., 20
  mm longer than in the other segments. Besides these fluctuations,
  the stability plot of the multi-segment resonator is identical to
  the stability plot of the single-segment resonator.  The red curve
  (1 segment containing 16 passes) represents the stability plot for a
  4f-based multi-pass resonator. Its stability region is 8 times
  smaller than the reference because it shrinks with the number of
  passes.  
  The blue curve (2 segments, each
  containing 8 passes) represents a merged resonator concept (see
  Sec.~\ref{sec:merging}) having two segments
  containing 4f-imaging stages.  (b) Corresponding average (over a
  round-trip) transmission through the Gaussian aperture at the
  thin-disk for variations of the thin-disk thermal lens.  }
\end{figure}

This behavior can be generalized to many segments: the stability
region, mode waist and losses per pass (averaged over a round-trip)
for the 8-segment multi-pass resonator of Fig.~\ref{fig:2}~(d) with
small segment-to-segment deviations turn out to be practically
identical with the stability region, the mode waist and losses per
pass of the underlying segment as demonstrated in Fig.~\ref{fig:5}
(compare green solid  with dashed black curves).
The same figure for comparison also shows the smaller stability range
featured by the multi-pass resonator based on 4f-imaging stages having
the same number of passes and beam size at the active medium.

In summary, the soft aperture effects occurring naturally in the
pumped active medium grant the realization of a multi-pass oscillator
as a concatenation of several nearly-identical optical segments.
When considering aperture effects, the stability region and losses per
pass of our multi-segment resonator is practically identical to the
stability region of the single-segment resonator.
Therefore contrarily to the 4f-based multi-pass resonator, the
stability region of our multi-pass resonator does not shrink with the
number of passes.

\section{Proof of principle}
\label{sec:proof}
\begin{figure}[t!]
\centering\includegraphics[width=1\textwidth]{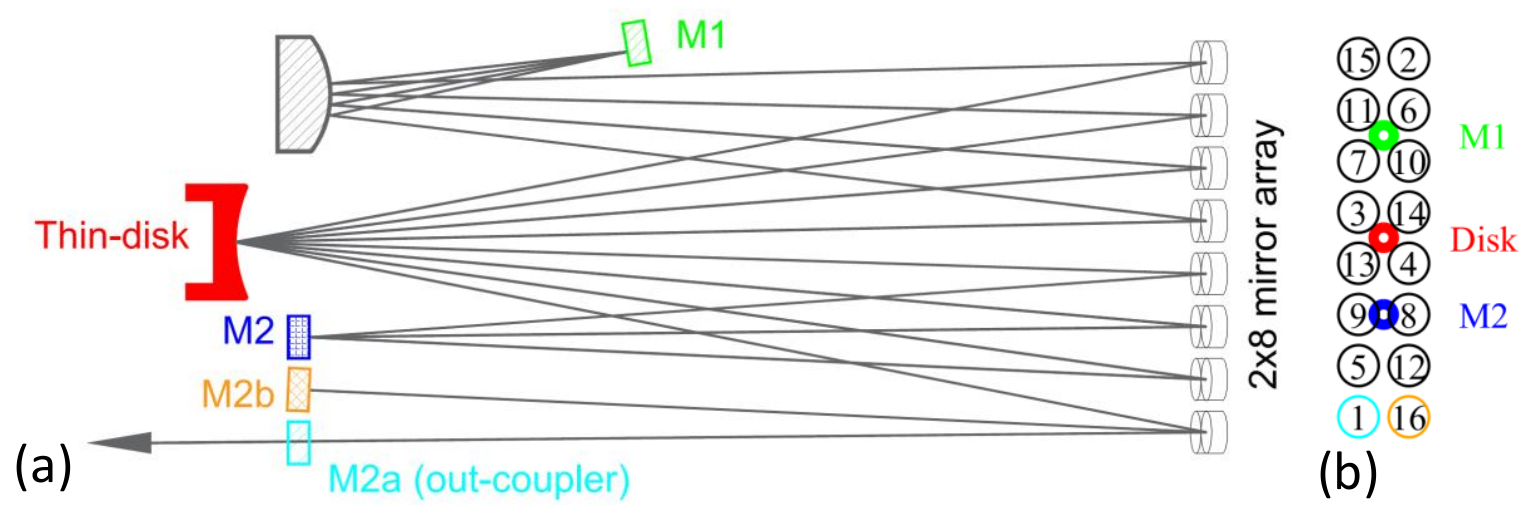}
\caption{\label{fig:6} (Color online) (a) Possible realization of a
  multi-pass oscillator with 16 reflections at the thin-disk per
  round-trip achieved by concatenating 8 identical segments.  The beam
  routing requires a mirror-array of 16 flat mirrors which can be
  individually adjusted. One of the two end-mirrors M2a and M2b could be used
  as an out-coupler. (b) Mirror-array working principle. The
  beam routing at the mirror-array plane follows the given numbering
  and is achieved by successive reflections at the thin-disk, mirror
  M1 and mirror(s) M2. The projection of these elements are
  indicated. }
\end{figure}
For a proof of principle we transformed the multi-pass
amplifier~\cite{Antognini2009, Schuhmann2015} that we developed for
spectroscopy of muonic atoms~\cite{Pohl2010, Antognini2013} into a
multi-pass oscillator by adding two end-mirrors.
This design schematically depicted in Fig.~\ref{fig:6} fulfills our
requirements of sufficiently small segment-to-segment variations as it
uses the same thin-disk and the same convex mirror in all segments.
Moreover the mirror array that is used to fold the beam providing 
several passes on the same thin-disk also guarantees similar path
lengths and small incident angles.

The beam routing in this multi-pass oscillator obeys the following
scheme.
Starting from the out-coupler M2a the beam is reflected at the
array-mirror 1 towards the thin-disk. From here, it proceeds towards
the array-mirror 2 and the convex mirror until it reaches M1. From M1
the beam travels back to the array at array-mirror 3, then to the
thin-disk and the array-mirror 4 until it reaches the mirror M2.
This scheme is iterated until the beam passes the array-mirror 16 and
is back-reflected at the second end-mirror M2b. From here the beam
propagates the same path  backwards  until it reaches again
mirror M2a closing the round-trip.
The beam routing at the mirror-array position given by the numbering
as shown in Fig.~\ref{fig:6} (b) can be thus understood as alternating
point-reflections at the thin-disk, M1 and M2 mirrors projections.

\begin{figure}[t!]
\centering \includegraphics[width=0.65\textwidth]{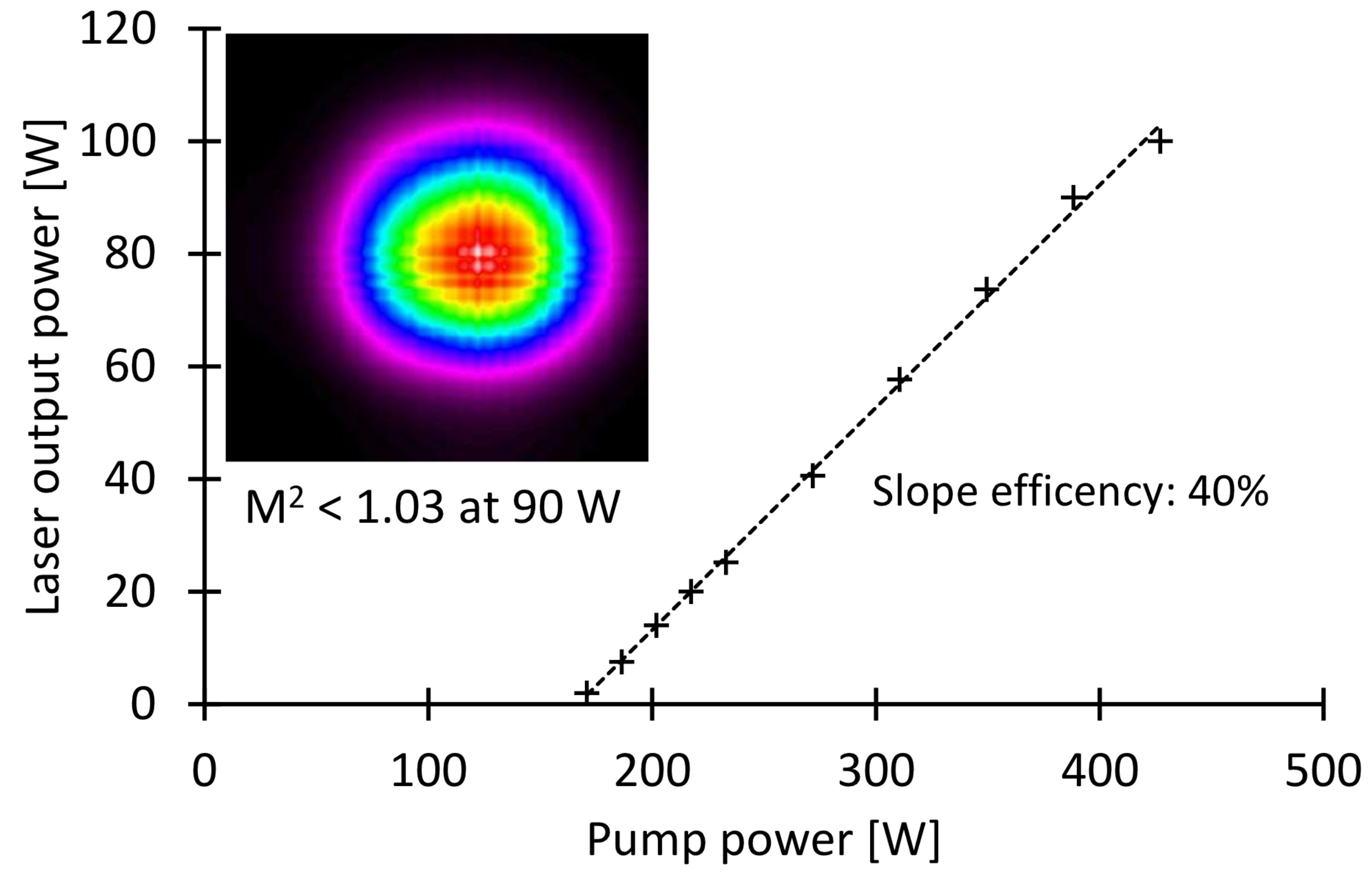}
\caption{\label{fig:7} (Color online) Input-output characteristics of
  the multi-pass resonator based on our design with 16 reflections per
  round-trip. The measurements have been accomplished in a cw
  operation for a Yb:YAG thin-disk of 345~$\mu$m thickness, a 940~nm pump
  wavelength and an out-coupling mirror reflectivity of 50\%. The
  inset shows the measured output beam.  }
\end{figure}
The multi-pass oscillator (with 16 reflections at the
thin-disk per round-trip) whose underlying segment specifications are given in 
Fig.~\ref{fig:2} (a) has been tested in cw mode using a flat
out-coupler with 50\% transmission.
As active medium, a 345~$\mu$m thick Yb:YAG thin-disk with 5\% nominal
doping concentration contacted by TRUMPF to a water-cooled CVD-diamond
heat sink having a 4~m radius of curvature has been used.
Even though the choice of the thin-disk parameters were optimized for low
repetition rate Q-switched operation, encouraging output powers and 
slope efficiency (40\% in fundamental mode operation) have been
observed as shown in Fig.~\ref{fig:7}.
This represents the first preliminary demonstration of the
applicability of our multi-pass oscillator concept, in particular
showing that the soft aperture effects naturally present in the
pumped active medium are sufficient to suppress the instabilities
related to the various segment-to-segment asymmetries associated with
the practical realization of a multi-pass scheme.

\section{A simple way to improve the multi-pass resonator based on 4f imaging}
\label{sec:merging}

\begin{figure}[t!]
\centering\includegraphics[width=1\textwidth]{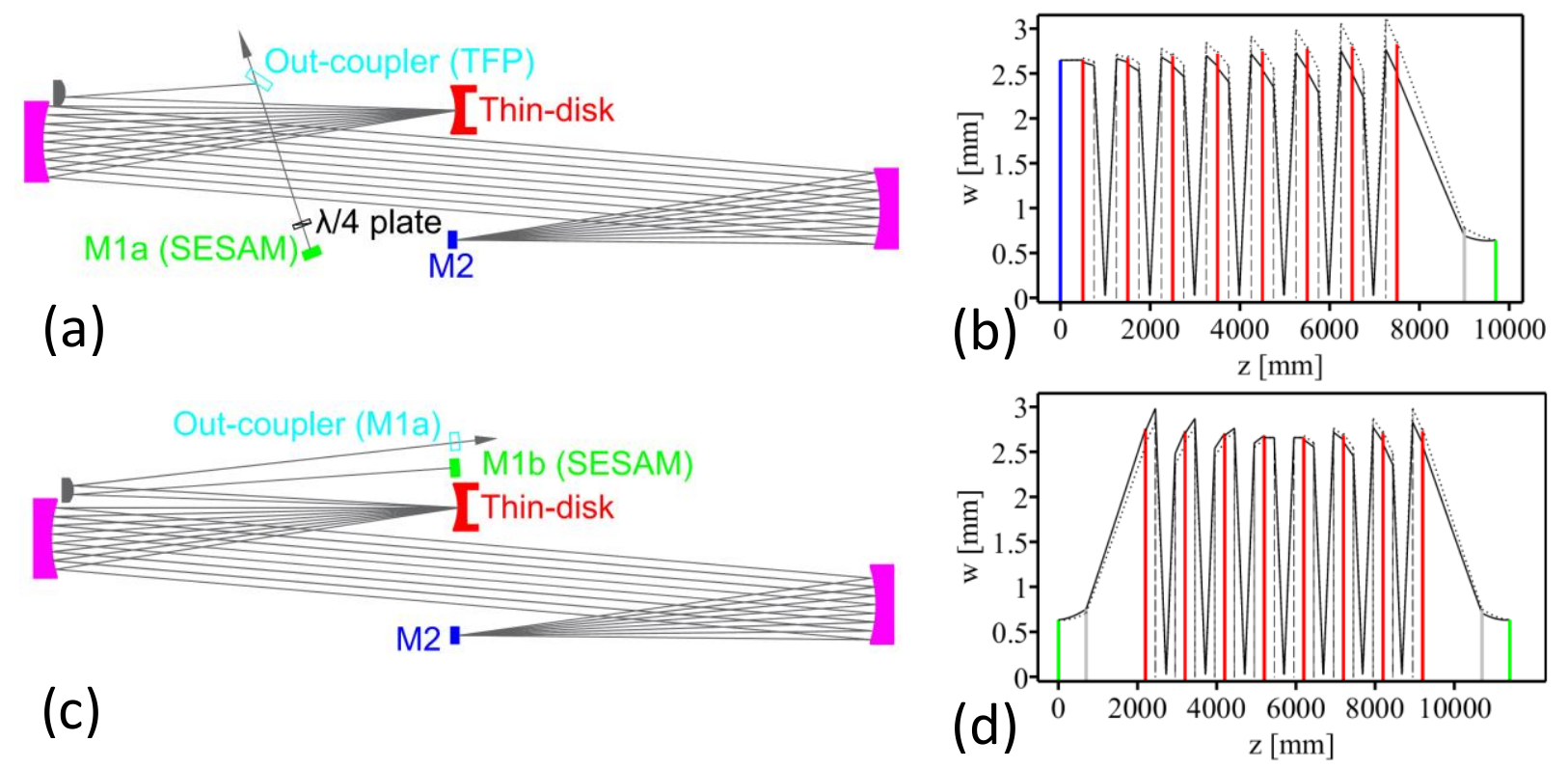}
\caption{\label{fig:8} (Color online) (a) Multi-pass oscillator based
  on 4f-imaging stages. Multiple 4f-imaging stages can be implemented
  using the same optical elements which simplifies the mechanical
  realization and decreases the production costs. TFP: thin-film
  polarizer. (b) Corresponding eigen-mode waist evolution.  Due to
  soft aperture effects the back (dotted line) and forth (continuous
  line) propagations have different waists. The vertical red lines
  represent the position of the thin-disk.  (c) Schematic of the
  merged concept with two segments, each containing half the number of
  reflections on the thin-disk as in (a). (d) Corresponding eigen-mode waist evolution. }
\end{figure}

The multi-pass oscillators based on  4f-imaging stages
show an enhanced sensitivity to thermal lens effects as illustrated by
the shrinking of the stability region with the number of passes in
Fig.~\ref{fig:1}.
However, a major advantage of the 4f-scheme is that a sequence of
several 4f-imaging stages can be realized using only few optical
elements as shown in Fig.~\ref{fig:8}~(a) whose working principle is
detailed in~\cite{Kumkar2003, Lundquist2013, Schuhmann2015a}.
On the other hand, our multi-pass oscillator concept has a superior
stability for variations of the thermal lens, but it requires an array
of mirrors resulting in increased mechanical complexity.

In Fig.~\ref{fig:8}~(b) we present an optical layout which results
from merging the two concepts.  
It consists of a  concatenation of two optically stable segments
(according to our scheme) each containing a multi-pass sequence based
on 4f-imaging stages.
As this merging can be achieved by a simple rearrangement of the
optics used in the 4f-based system, it inherits a similar beam waist
evolution (see Fig.~\ref{fig:8} (b) and (d)) and its simplicity,
qualifying this scheme for industrial applications.
At the same time this merged layout shows improved stability because
the stability region of our concept does not shrink with the number of
segments.
The resulting stability region of this merged scheme (two segments,
each containing $N/2$ 4f-propagations) is a factor of two larger
compared to the standard 4f-schemes (one segment with $N$
4f-propagations) as visible from the comparison of the blue and red
curves in Fig.~\ref{fig:5}, provided both multi-pass resonators have
the same number of passes.
Thus, with a simple rearrangement of the beam path structure of the
standard 4f-design (but using the same optical elements) a factor of
two larger stability region can be obtained opening the way to larger
pump power density, beam waists and number of passes.

\section{Conclusions}
We have presented a multi-pass resonator scheme as a sequence of
nearly-identical optically stable segments, each segment containing
the same active medium featuring soft-aperture effects.
The stability region of such a multi-segment resonator does not
decrease with the number of segments.
Therefore this concept solves the limitations of state-of-the-art
multi-pass resonators based on 4f-imaging stages which feature a
shrinking of the stability region with the number of passes on the
active medium.

We have demonstrated that it is essential to include the soft aperture
effects occurring in the active medium into the simulations.
They
suppress the formation of gaps within the stability region which would
arise as a consequence of small segment-to-segment asymmetries
associated with the practical realization of a multi-pass system.
This has been achieved by implementing effective Gaussian apertures
into the ABCD-matrix formalism as lenses with
imaginary focal length.

The multi-pass resonator concept presented here requires small
segment-to-segment asymmetries  achievable using the same
active medium in all passes.
Larger segment-to-segment deviations would cause increased losses
(decreased transmission through the aperture) which strongly reduce
laser efficiency or even disrupt laser operation.

This multi-pass resonator layout is particularly suited for ultrafast
lasers where the mode-locking mechanism is based on SESAM
technologies~\cite{Saraceno2015}.
The SESAM could be placed at the position of one of
the resonator end-mirrors (e.g. mirror M2b in Fig.~\ref{fig:6})  as
at this position there is minimal intracavity intensity and the SESAM
mirror would be intersected only once per round-trip.
This scheme having several passes on the active medium and large
cavity lengths paves the way for energy and power scaling of
mode-locked lasers expanding greatly the range of applications for 
ultrashort pulses delivered directly by a laser oscillator.

\section{Acknowledgments}
We would like to thank R. Pohl and F. Kottmann. We acknowledge the
support from the Swiss National Science Foundation: Projects
SNF\_200020\_159755 and SNF\_200021\_165854.

\end{document}